\documentclass[twocolumn]{aastex631}

\newcommand{\C}{$^{12}$C}

\newcommand{\Fe}{$^{56}$Fe}

\begin{document}

\title{Diffusion Coefficients of $^{56}$Fe in C-O and O-Ne White Dwarfs}

\author{Matthew E. Caplan}
\affiliation{Department of Physics, Illinois State University \\ Normal, IL 67190, USA}
\email{mecapl1@ilstu.edu}



\begin{abstract}

The diffusion coefficients of neutron rich nuclei in crystallizing white dwarf (WD) stars are essential microphysics input for modeling the evolution of the composition profile. Recently, molecular dynamics simulations have been used to compute diffusion coefficients for realistic mixtures of C-O and O-Ne WDs with many trace nuclides that could be important sedimentary heat sources such as $^{22}$Ne, $^{23}$Na, $^{25}$Mg, and  $^{27}$Mg. In this brief note, I repeat these simulations but now include $^{56}$Fe. I find that for the large charge ratios involved in these mixtures the empirical law developed in our earlier work tends to under-predict diffusion coefficients in the moderately coupled regime by 30 to 40 percent. As this formalism is presently implemented in the stellar evolution code MESA, it is important for authors studying mixtures containing heavy nuclides like $^{56}$Fe to be aware of these systematics. However, the impact on astrophysics is expected to be small. 


\end{abstract}



\section{Introduction} \label{sec:intro}

\begin{figure*}
\centering  
\includegraphics[trim=15 130 180 0,clip,width=0.48\textwidth]{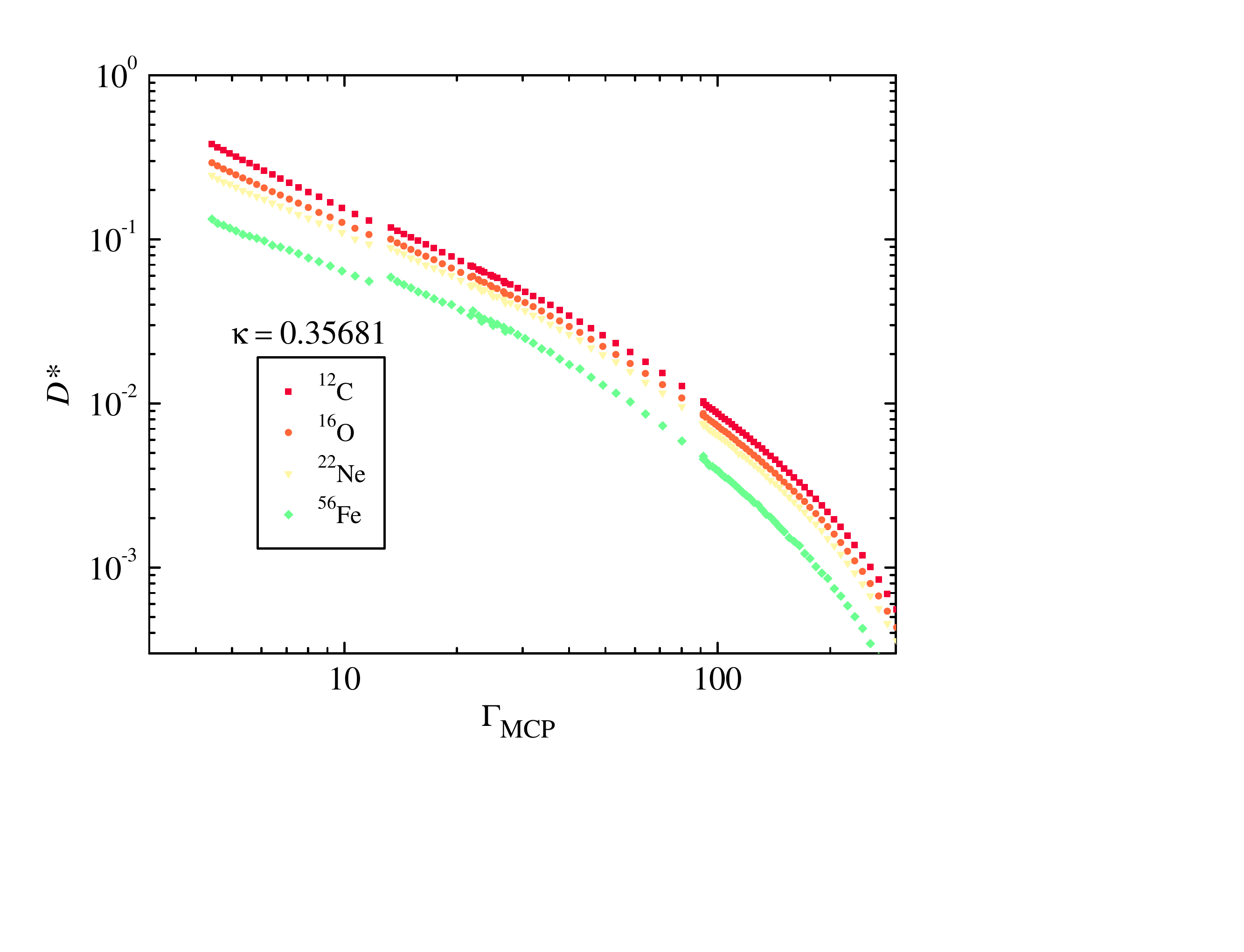}
\includegraphics[trim=15 130 180 0,clip,width=0.48\textwidth]{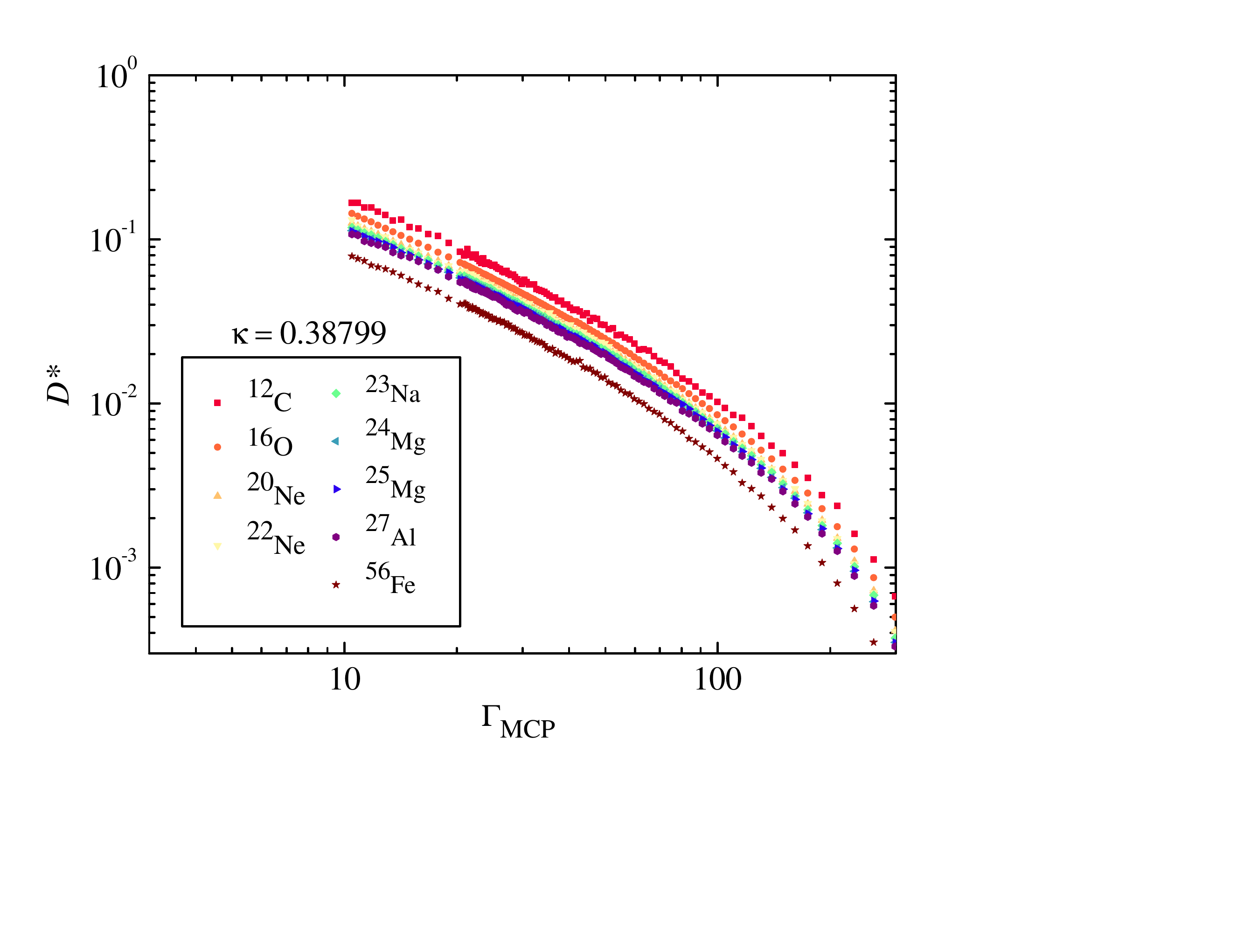}
\includegraphics[trim=25 130 210 10,clip,width=0.48\textwidth]{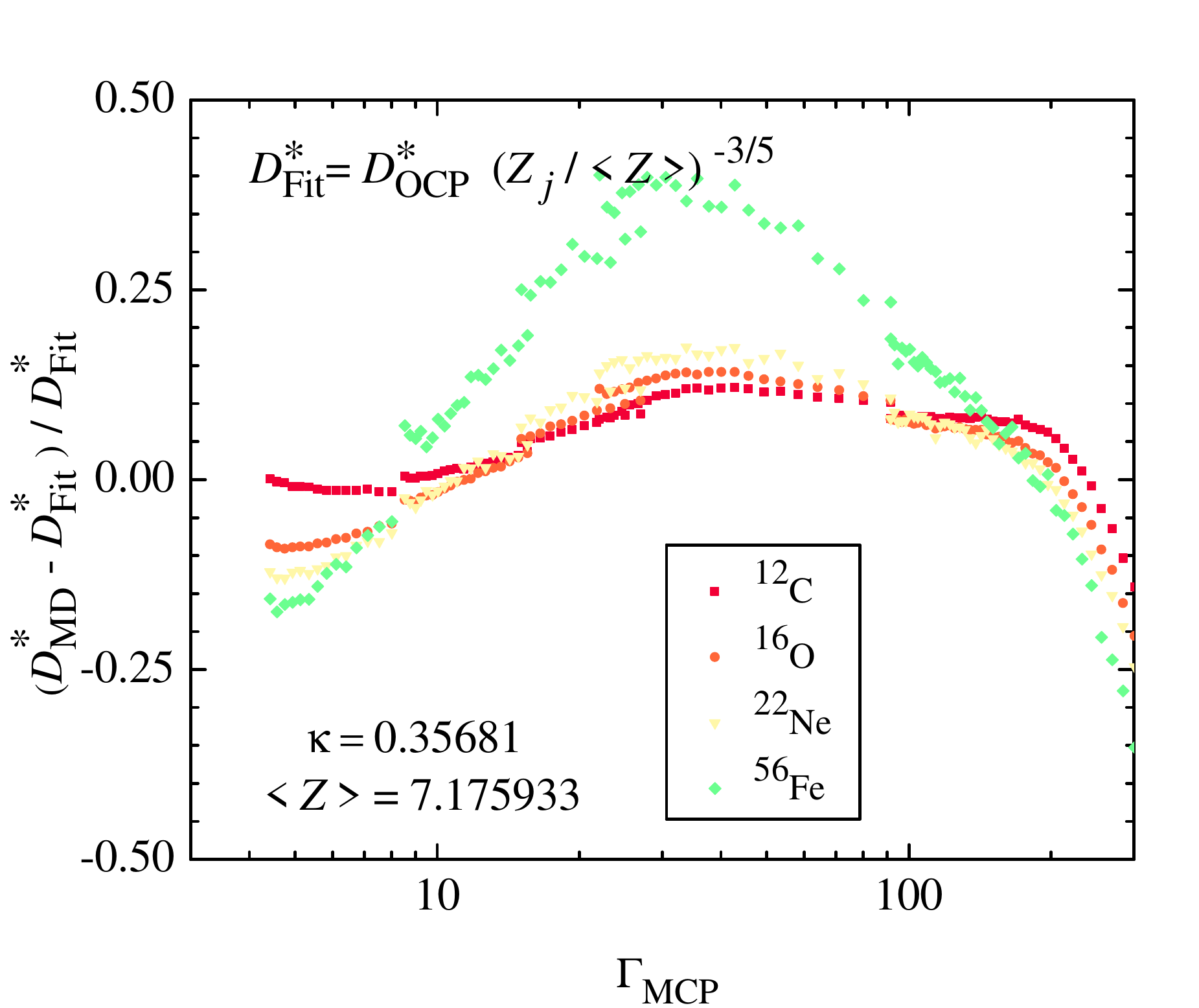}
\includegraphics[trim=25 130 210 10,clip,width=0.48\textwidth]{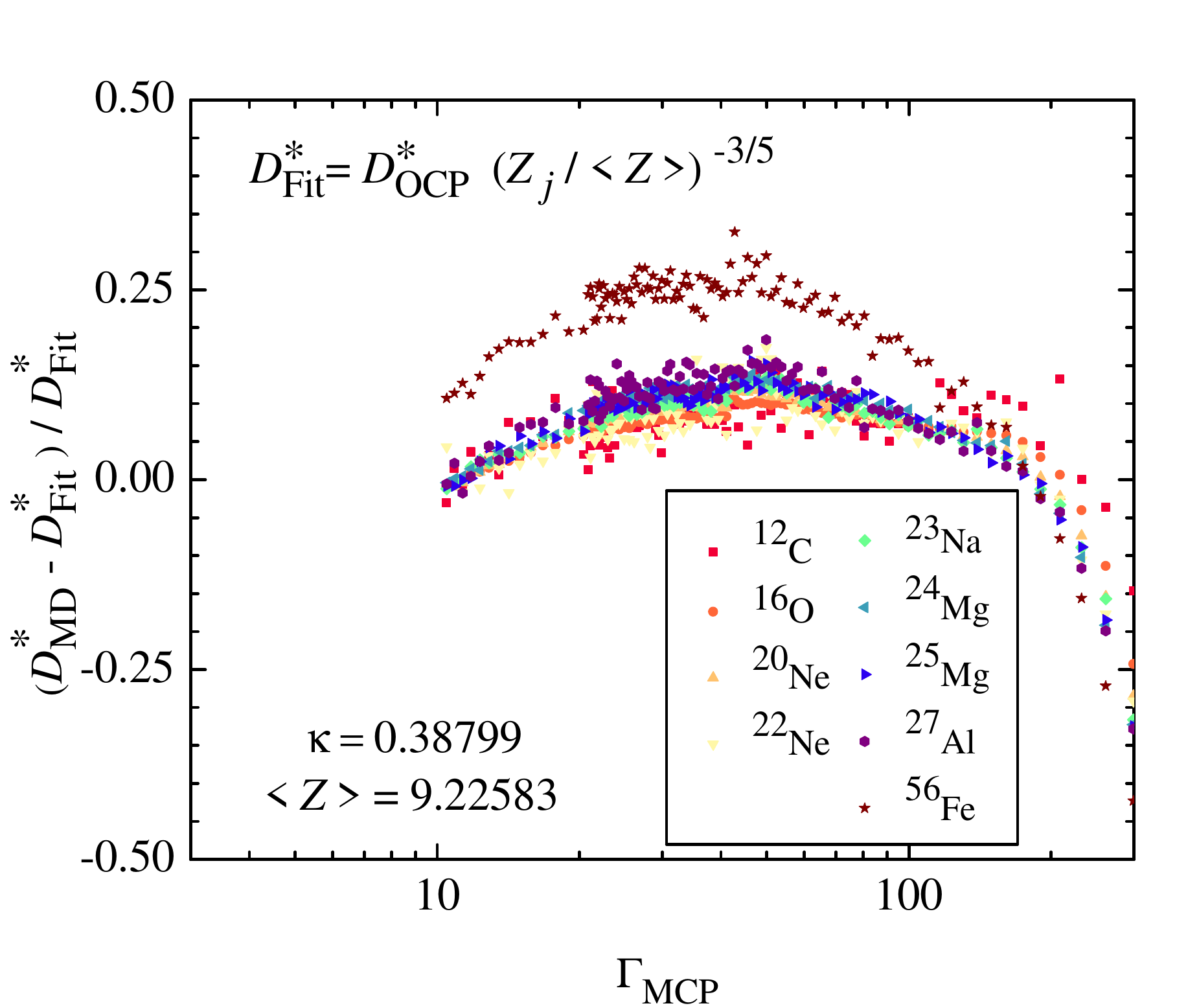}
\caption{ (Top) Dimensionless diffusion coefficients for nuclei in our simulations and (bottom) residuals between the MD and the proposed fits. While $D^*_{\rm Fe}$ is over-predicted for both (left) C-O and (right) O-Ne mixtures, I find that the fit is marginally better for the O-Ne mixture due to the smaller charge ratio between the \Fe\ nuclei and the background. }
\label{fig:MD} 
\end{figure*} 

Recent work has calculated the diffusion coefficients of nuclei in complex mixtures  for white dwarf (WD) modeling. Using molecular dynamics (MD) simulations, \cite{caplan2022} was able to develop a physically motivated law for diffusion in pure systems and generalize that law to mixtures.\footnote{Readers are encouraged to familiarize themselves with \cite{caplan2022}.}

For the one-component plasma (OCP) interacting through a screened Coulomb potential $U(r) = (e^2 Z^2/r)\mathrm{e}^{-\kappa r}$, the following law was presented for the dimensionless $(D^* = D/\omega_p a^2)$ diffusion coefficient

\begin{equation}\label{eq:Dmodel}
    D_{\rm OCP}^*(\Gamma,\kappa) =  \sqrt{\frac \pi 3} \frac{ A(\kappa) \Gamma^{-5/2}}{ \ln \left( 1 + \frac{C(\kappa) }{ \sqrt{3}} \Gamma^{-3/2} \right) } e^{-B(\kappa) \Gamma}~,
\end{equation}

\noindent with $\Gamma$ the plasma coupling strength ($\Gamma \propto 1/T$) and $\kappa$ the screening length, while A, B, and C are obtained from the parametrization found in \cite{caplan2022}. That work considered $0 \lesssim \kappa \lesssim 1$, making this law suitable for both white dwarf interiors and neutron star crusts \citep[e.g. Fig. 3 in ][]{Blouin2021a}.

This was generalized to mixtures by arguing that each species in a mixture should experience Stokes-Einstein drift because the mixture viscosity is a global property \citep{Bauer2020}. As such, there is only parameter for each species which is the effective ion radius $R$, with $D^* \propto R^{-1}$. This radius is then taken to be an effective charge radius dependent on the species charge $Z_j$.

The diffusion coefficient of a species with charge $Z_j$ in a mixture was found to be well fit by treating the background as an equivalent one-component liquid of charge $\langle Z \rangle$, the mixture average charge. In this formalism, every nucleus is treated as an impurity moving through a pure fluid that is expected to have the same effective viscosity as the mixture. For an arbitrary multi-component plasma (MCP), the diffusion coefficients of the species-wise components are calculated by

\begin{equation}\label{eq:Dmodel2} 
    D_j^* = D_{\rm OCP}^* \left( \frac{Z_j}{ \langle Z \rangle} \right)^{-0.6}~,
\end{equation}

\noindent where  $D_{\rm OCP}^*(\Gamma_{\rm MCP},\kappa)$ is the equivalent OCP to the mixture at a given temperature (or mixture coupling $\Gamma_{\rm MCP}$) with the same screening length. The power $-0.6$ is empirically determined from the MD and likely depends strongly on the screening length $\kappa$ if considering mixtures with $\kappa > 1$. 


\cite{caplan2022} only considered a small number of astrophysically relevant neutron rich nuclei, specifically $^{22}$Ne, $^{23}$Na, $^{25}$Mg, and  $^{27}$Mg. Recently, papers computing phase diagrams of $^{56}$Fe in C-O \citep{Caplan2021iron} and O-Ne \citep{caplan2023iron} mixtures find that the large charge of \Fe\ (relative to the background) can result in strong separation at early times producing substructures in the WD that are strongly enhanced in \Fe\, such as Fe-rich shells and inner cores. 
As the diffusion formalism of \cite{caplan2022} is now implemented the in the stellar evolution code MESA, it is important to validate Eqs. \ref{eq:Dmodel} and \ref{eq:Dmodel2} for mixtures with larger charge ratios \citep{MESA2023}. I am therefore motivated to revisit the mixtures from \cite{caplan2022} and include Fe.

\section{Results} \label{sec:res}

The MD formalism used here is the same as \cite{caplan2022} and is described in detail in past work. In summary, nuclei are treated as point particles interacting through a screened Coulomb potential and evolved using the velocity Verlet algorithm. Diffusion coefficients are calculated from the mean squared displacement of nuclei over an appropriately chosen time interval. 

The mixtures contain 65536 nuclei and are nearly identical to those in \cite{caplan2022} except now 400 nuclei from the C-O or O-Ne background are replaced with \Fe. This is effectively in the impurity limit so I do not expect a major impact on the other components of the mixture. I also simulate at a smaller range of $\Gamma$ and use an ensemble of 25 simulations rather than the 40 in \cite{caplan2022}. For the C-O mixture four sets of simulations are separately run, each covering a factor of a few in $\Gamma_{\rm MCP}$. For the O-Ne mixture only three sets are run, omitting $\Gamma_{\rm MCP} \lesssim 10$. 

The top of Fig. \ref{fig:MD} shows $D^*$ for every species in both mixtures and below them are the residuals with respect to Eq. \ref{eq:Dmodel2}. 
In the C-O mixture (left) one can see that the \Fe\ continues the trend toward lower $D^*$ at higher charges, but the residual (bottom) with respect to the law is quite large, reaching a maximum near 40 percent around $\Gamma_{\rm MCP} \approx 30$. The O-Ne mixture (right) shows the same trend but performs better, likely due to the smaller charge ratio, with peak errors closer to 30 percent. Taken together, these suggest that our law systematically underpredicts the diffusion coefficient of large charge impurities moving through a low charge background at intermediate coupling strengths. For $\Gamma_{\rm MCP} \gtrsim 200$ one also observes an apparent overprediction in the fit as the residuals very sharply decrease. However, this is almost certainly due to supercooling in the MD as the mixtures would crystallize here if run sufficiently long for a seed crystal to spontaneously nucleate. 

\section{Discussion} \label{sec:disc}

In summary, the law for mixtures in Eq. \ref{eq:Dmodel2} underpredicts for $10 \lesssim \Gamma_{\rm MCP} \lesssim 100$ when considering trace nuclei with large charge ratios relative to the background. While some small systematic spread could be observed in the mixtures previously studied, the largest $(Z_j / \langle Z \rangle)$ considered was only approximately 1.4. When including \Fe, with $(Z_j / \langle Z \rangle) \approx 4$ in C-O, it is immediately apparent that the underprediction is systematic with charge ratio.

While Eq. \ref{eq:Dmodel2} has errors as large as 40 percent for \Fe\ in C-O and 30 per cent for \Fe\ in O-Ne, this might not have too large of an affect on astrophysics modeling. Sedimentation and separation processes depend most strongly on $D^*$ near crystallization. As WDs rapidly cool, most time is spent at low temperatures so the astrophysics should be most sensitive to $100 \lesssim  \Gamma_{\rm MCP} \lesssim 200$ and the law shows good performance here.

Improvements to Eq. \ref{eq:Dmodel2} to correct for this systematic are not immediately obvious. Refitting the -0.6 exponent is unsuitable, as that would systematically shift $D^*_j$ of all species at all $\Gamma$. The underlying fits for the OCP might be improved, but this will likely only result in a marginal improvement. It should perhaps be recognized that the choice of $(Z_j  / \langle Z \rangle )^{-0.6}$ was somewhat arbitrary, and could be better motivated physically. One could also imagine fitting this residual empirically with a lognormal distribution, as in \cite{caplan2021mnras}, but this will have limited utility. Developing a general form for mixtures would likely require a larger survey of binary (and ternary) MD mixtures for many charge ratios and screening lengths. It is not immediately apparent if this is worthwhile, as the current laws represent a significant improvement on the \cite{Stanton2016} diffusion coefficients that were previously implemented in MESA and are likely sufficiently precise for present purposes in astrophysics \citep{MESA2023}. 

I conclude by emphasizing that these large residuals are not unique to $^{56}$Fe, but are rather a property of the large charge ratio of $^{56}$Fe with respect to the background in the specific simulations reported on here. For a mixture of nearly pure $^{56}$Fe (or similar nuclei) that one might find in a neutron star, I expect Eqs. \ref{eq:Dmodel} and \ref{eq:Dmodel2} to perform quite well.

\begin{acknowledgments}
I thank Simon Blouin and Evan Bauer for their thoughtful feedback. This research was supported in part by the National Science Foundation under Grant No. NSF PHY-1748958. M.C. acknowledges support as KITP Scholar.
\end{acknowledgments}

\providecommand{\noopsort}[1]{}\providecommand{\singleletter}[1]{#1}%




\begin{thebibliography}{}
\expandafter\ifx\csname natexlab\endcsname\relax\def\natexlab#1{#1}\fi
\providecommand{\url}[1]{\href{#1}{#1}}
\providecommand{\dodoi}[1]{doi:~\href{http://doi.org/#1}{\nolinkurl{#1}}}
\providecommand{\doeprint}[1]{\href{http://ascl.net/#1}{\nolinkurl{http://ascl.net/#1}}}
\providecommand{\doarXiv}[1]{\href{https://arxiv.org/abs/#1}{\nolinkurl{https://arxiv.org/abs/#1}}}

\bibitem[{{Bauer} {et~al.}(2020){Bauer}, {Schwab}, {Bildsten}, \&
  {Cheng}}]{Bauer2020}
{Bauer}, E.~B., {Schwab}, J., {Bildsten}, L., \& {Cheng}, S. 2020, \apj, 902,
  93, \dodoi{10.3847/1538-4357/abb5a5}

\bibitem[{{Blouin} \& {Daligault}(2021)}]{Blouin2021a}
{Blouin}, S., \& {Daligault}, J. 2021, \pre, 103, 043204,
  \dodoi{10.1103/PhysRevE.103.043204}

\bibitem[{Caplan \& Freeman(2021)}]{caplan2021mnras}
Caplan, M., \& Freeman, I. 2021, \mnras, 505, 45

\bibitem[{{Caplan} {et~al.}(2022){Caplan}, {Bauer}, \& {Freeman}}]{caplan2022}
{Caplan}, M.~E., {Bauer}, E.~B., \& {Freeman}, I.~F. 2022, \mnras, 513, L52,
  \dodoi{10.1093/mnrasl/slac032}

\bibitem[{{Caplan} {et~al.}(2023){Caplan}, {Blouin}, \&
  {Freeman}}]{caplan2023iron}
{Caplan}, M.~E., {Blouin}, S., \& {Freeman}, I.~F. 2023, \apj, 946, 78,
  \dodoi{10.3847/1538-4357/acbfaa}

\bibitem[{{Caplan} {et~al.}(2021){Caplan}, {Freeman}, {Horowitz}, {Cumming}, \&
  {Bellinger}}]{Caplan2021iron}
{Caplan}, M.~E., {Freeman}, I.~F., {Horowitz}, C.~J., {Cumming}, A., \&
  {Bellinger}, E.~P. 2021, \apjl, 919, L12, \dodoi{10.3847/2041-8213/ac1f99}

\bibitem[{{Jermyn} {et~al.}(2023){Jermyn}, {Bauer}, {Schwab}, {Farmer}, {Ball},
  {Bellinger}, {Dotter}, {Joyce}, {Marchant}, {Mombarg}, {Wolf}, {Sunny Wong},
  {Cinquegrana}, {Farrell}, {Smolec}, {Thoul}, {Cantiello}, {Herwig}, {Toloza},
  {Bildsten}, {Townsend}, \& {Timmes}}]{MESA2023}
{Jermyn}, A.~S., {Bauer}, E.~B., {Schwab}, J., {et~al.} 2023, \apjs, 265, 15,
  \dodoi{10.3847/1538-4365/acae8d}

\bibitem[{{Stanton} \& {Murillo}(2016)}]{Stanton2016}
{Stanton}, L.~G., \& {Murillo}, M.~S. 2016, \pre, 93, 043203,
  \dodoi{10.1103/PhysRevE.93.043203}

\end{thebibliography}
\end{document}